\documentclass[12pt]{article}
\usepackage{amsmath,amssymb}
\usepackage{eucal}

\usepackage[dviwindo]{graphicx}
\newcommand{\bs}{\begin{subequations}}
\newcommand{\es}{\end{subequations}}
\numberwithin{equation}{section}
\newcommand{\ben}{\begin{eqnarray}}
\newcommand{\een}{\end{eqnarray}}
\newcommand{\la}{\label}
\begin{document}

\title{On the Exact Solutions of the Regge-Wheeler Equation in the Schwarzschild
Black Hole Interior}
\vskip 1.5truecm
\author{Plamen~P.~Fiziev\thanks{Department of Theoretical Physics, University of
Sofia, Boulevard 5 James Bourchier, Sofia 1164, Bulgaria;
E-mail:\,\,\,fiziev@phys.uni-sofia.bg}}
\date{}
\maketitle
\begin{abstract}
We solve the Regge-Wheeler equation for axial perturbations of the
Schwarzschild metric in the black hole interior in terms of Heun's
functions and give a description of the spectrum and the
eigenfunctions of the interior problem.  The phenomenon of
attraction and repulsion of the discrete eigenvalues of
gravitational waves is discovered.
\end{abstract}
\vskip .5truecm
{\bf Keywords:}  Schwarzschild black hole, gravitational waves,
Regge-Wheeler equation, Heun's functions.
\vskip .5truecm
{\bf PACS:} 04.70.Bw, 04.30.-w., 04.30 Nx

% main text

\sloppy

\section{Introduction}

The perturbations of the gravitational field of black holes (BH) is
a very active and large area of analytical, numerical, experimental
and astrophysical research. Ongoing and future experiments, based on
perturbation analysis and numerical calculations, are expected to
give critical tests of the existing theories of gravity
\cite{KipTorn03, GW}. For the experimental study of the
gravitational waves a unprecedented precision of the already
existing and future detectors must be reached \cite{KipTorn03, GW}.
Since this is an extremely arduous and expensive task, an important
goal of the theory is to reach detailed description of all features
of the studied phenomena.

For the study of the perturbations outside the BH horizon one uses
the quasi-normal modes (QNM) with well known complex spectra
\cite{QNM}. QNM do not form a complete basis of functions in the
space of the linear metric's perturbations. Nevertheless, they are
widely used as a tool for theoretical study of the perturbations in
the outer domain of BH. The QNM are believed to be relevant to the
quantum gravity, too \cite{QNMQG}.

Analogous tool for the study of perturbations of the black hole
interior is not known even in the simplest case of Schwarzschild
black holes (SBH). Since the BH-spacetimes are considered to
describe uniform physical objects with their exterior and interior,
separated by an event horizon, it is natural to study the
perturbations of the gravitational field in both domains, at least
by reason of completeness.

In the classical theory of BH it seems reasonable to exclude from
the most of the considerations the inner domain, because it does not
influence {\em directly} the observable outer domain. Nevertheless,
the excitations of degrees of freedom in the inner domain may be
essential for the consideration of the total BH entropy and for
construction of the quantum theory of BH. The interior excitations
may yield observable effects in the exterior of the BH, due to the
quantum effects, see, for example, the recent article \cite{BMR} and
references therein. To discuss such problems one needs a proper
mathematical description of the BH-interior's perturbations.

Another reason to study the BH interior was discovered in the very
recent article \cite{LBLR}: It was shown that without excision of
the inner domain around the singularity one can improve dramatically
the long-term stability of the numerical calculations. Although the
region of the space-time that is causally disconnected can be
ignored for signals and perturbations traveling at physical speeds,
numerical signals, such as gauge waves or constraint violations, may
travel at velocities larger then that of light and thus leave the
physically disconnected region.

Thus the study of the perturbations of the BH interior becomes an
important issue both from analytical and numerical point of view.

In the present article we are starting the analytical and numerical
investigation of the mathematical problem, introducing a large set
of solutions for description of the perturbation of the SBH interior
and studying the basic properties of the corresponding functions. We
believe that this will help a more deep understanding of the
problems of BH physics.

We show that one can find some unexpected new results for the
perturbations of the SBH interior. In particular, we have discovered
that variations of the boundary conditions at the event horizon
$r=1$ produce attraction and repulsion of the discrete spectral
levels of the gravitational waves in the SBH interior.

\section{General Consideration of the Interior Problem}
The proper mathematical ground for the investigation of the problem
is the Regge-Wheeler equation (RWE) \cite{RW}:
\ben  \partial_t^2 \Phi_{s,l} +\left(-\partial^2_{x}
+V_{s,l}\right)\Phi_{s,l} = 0. \la{RW}\een
Its study has a long history as well as important and significant
achievements \cite{QNM, RW}.

The RWE describes the axial perturbations of the Schwarzschild
metric of mass $M$ in the first order linear approximation. In Eq.
(\ref{RW})
$x\!=\!r\!+\!r_{\!{}_{Sch}}\ln\left(|r/r_{\!{}_{Sch}}\!-\!1|\right)$
denotes the Regge-Wheeler "tortoise" coordinate, $r_{\!{}_{Sch}}=2M$
is the Schwarzschild radius and $\Phi_{s,l}(t,r)$ is the radial
function for perturbations of spin $s$ and angular momentum
$l\!\geq\!s$. Hereafter we use units $r_{\!{}_{Sch}}\!=\!2M\!=\!1$.
The effective potential in RWE (\ref{RW}) reads
$$V_{s,l}(r)=\left(1-{{1}\over r}\right)\left({{l(l+1)}\over {r^2}}
+{{1-s^2}\over {r^3}} \right).$$  For  $r\in[0, 1]$ the area radius
$r$ in it can be expressed explicitly as a function of the variable
$x$ using Lambert W-function \cite{W}: $r=W(-e^{x-1})+1$.

The most important case $s\!=\!2$ describes the gravitational waves
and is our main subject here.

The standard ansatz $\Phi_{s,l}(t,r)=R_{\varepsilon,s,l}(r)
e^{i\varepsilon t}$ brings us to the stationary problem in the outer
domain $r>1$:
\ben
\partial^2_x R_{\varepsilon,s,l}+\left(\varepsilon^2-V_{s,l}\right)
R_{\varepsilon,s,l}=0. \la{R}\een
Its {\em exact} solutions were described recently \cite{F2005} in
terms of the confluent Heun's functions $HeunC\!\left(\alpha, \beta,
\gamma, \delta,\eta,r \right)$ \cite{Heun}.

Here we introduce for the first time the exact solutions of the
Regge-Wheeler equation in the interior of the SBH. It is well known
\cite{GR} that because of the change of the signs of the components
$g_{tt}=-1/g_{rr}=1-1/r$ of the metric, in this domain the former
Schwarzschild-time variable $t$ plays the role of radial space
variable $r_{in}=t\in (0,\infty)$ and the area radius $r\in(1,0)$,
plays the role of time variable. As a result, the Regge-Wheeler
"tortoise" coordinate $x\in (-\infty,0)$ presents a specific time
coordinate in the inner domain. For the study of the solutions of
Eq. (\ref{RW}) in this domain it is useful to stretch the last
interval to the standard one by a further change of the
time-variable: $x\to t_{in}=x-1/x\in (-\infty,\infty)$.

These notes are important for the physical interpretation of the
mathematical results. In particular, it is natural to write down the
interior solutions of the RWE in the form:
$\Phi_{\varepsilon,s,l}^{in}(r_{in},t_{in})=e^{i\varepsilon r_{in}}
R_{\varepsilon,s,l}\left(r(t_{in})\right)$, where
\ben r(t_{in})\!=\!W\!\left(-exp\left( {{t_{in}}\over{2}}\!-\!1\!-\!
\sqrt{\left({{t_{in}}\over{2}}\right)^2\!+\!1}\,\right)
\right)\!\!+\!1.\hskip .3truecm\la{r_tin}\een
The dependence of this solution on the interior radial variable
$r_{in}$ is simple. Its dependence on the interior time $t_{in}$ is
governed by the RWE (\ref{R}) with interior-time-dependent potential
$V_{s,l}$. In despite of this unusual feature of the solutions in
the SBH interior, this way we obtain a basis of functions, which are
suitable for the study of the corresponding linear perturbations.

Without any additional conditions the differential equation
(\ref{R}) has a reach variety of solutions which may describe
different physical problems. For a correct formulation of a given
physical problem one must restrict the class of the solutions using
proper boundary conditions. Since the physics at the boundary
influences strongly the solutions of the problem, the boundary
conditions are a necessary ingredient of the theory and define the
physical problem under consideration.

The solutions of the Eq. (\ref{R}), which are not regular at the
point $r=0$ are not stable, since they grow up infinitely as
$t_{in}\!\to\!+\infty$. In addition they have a complicated series
expansion with logarithmic terms \cite{Heun}. Hence, these solutions
are not single-valued functions and have no suitable properties for
physical applications\footnote{It is well known that analogous
solutions exist in the much more simple case of confluent
hypergeometric equation. For example, because of the same reasons
these are not allowed as solutions of the hydrogen atom problem in
quantum mechanics.}.

Solutions of the Eq. (\ref{R}), which grow up infinitely with
respect to the future direction of the exterior time $t$ exist in
the exterior domain of SBH, too. They are physically unacceptable
and therefore are excluded by formulation of a proper boundary
problem \cite{QNM, RW}. Only after that the physical problem is
fixed and one can prove the stability of the remaining solutions
(QNM) and small perturbations of general form with respect to the
exterior-time evolution \cite{NM}. All these solutions unavoidably
grow up infinitely both at the space infinity and at the horizon
\cite{QNM, RW, F2005}.

By analogy with the consideration in the outer domain of SBH, in the
inner one we are considering only the solutions, which are stable
with respect to the future direction of the interior time. Indeed,
the common physical sense states that only the stable solutions are
of physical interest. Therefore in this article we study the
solutions, which are regular at the point $r\!=\!0$, formulating the
corresponding boundary problem. The remaining solutions are stable
in the future direction of the interior time: $t_{in}\!\to\!+\infty$
and have a nontrivial spectrum with novel properties.

From a pure mathematical point of view in the present article we are
considering the two-singular-points-boundary problem \cite{Heun,
F2005} for the RWE on the interval $r\in[0,1]$. In this sense here
we present an extension of the article \cite{F2005}, where only
exact solutions of different types on the intervals $r\in
[0,\infty)$ and $r\in [r_*,\infty) \subset [0,\infty)$ were studied
in detail.

There exist one more possibility: to study
two-singular-points-boundary problem for the Eq. (\ref{R}) on the
interval $r\in [0,\infty)$ using some contour in the complex plane
${\cal C}_r$ with singular ends $0\,\,\hbox{and}\,\,\infty$, which
turns round on the third singular point $r=1$. Such solutions may
describe mutual perturbations both of SBH exterior and interior.
This interesting possibility seems to correspond to the
perturbations of the Kruscal-Szekeres extension of the Schwarzschild
solution \cite{GR}. It is beyond the scope of the present article
and will be considered somewhere else.

\section{The Exact Solutions in the SBH interior }
The RWE (\ref{R}) has three singular points: the origin $r\!=\!0$,
the horizon $r\!=\!1$ and the infinite point $r\!=\!\infty$. The
first two are regular singular points and can be treated on equal
terms. The last one is an irregular singular point, obtained by
confluence of two regular singular points in the general Heun's
equation. As a result, in the interior of the SBH around the regular
singular points we have the following local analytical solutions
\cite{F2005}\footnote{For a detailed explanation of the notations
and physical meaning of the solutions one can consult this article}:
\ben R_{\,\,\varepsilon,s,l}^{(0)-}(r)= r^{s+1} e^{ i\varepsilon
\left(\ln(1-r)+r\right)} HeunC\!\left(2i\varepsilon, 2s,
2i\varepsilon, 2\varepsilon^2,s^2-l(l+1),r \right), \la{r0}\een
\ben R_{\,\,\varepsilon,s,l}^{(1)\pm}(r)\!=\!r^{s\!+\!1} e^{\!\pm
i\varepsilon \left(\ln(1\!-\!r)\!-r\right)}
HeunC\!\left(\pm2i\varepsilon, \pm2i\varepsilon, 2s,
-2\varepsilon^2,2\varepsilon^2\!\!+\!s^2\!\!-\!l(l\!+\!1),1\!-\!r
\right)\!.\hskip 0.4truecm \la{r1}\een

The solution (\ref{r0}) is the only one, which is regular at the
point $r=0$. In addition it is the only single-valued local
solution, defined around this point.

The solution $R_{\,\,\varepsilon,s,l}^{(1)+}(r)$ in (\ref{r1}) can
be derived from the one, used in \cite{F2005}, applying proper
transformations of the Heun's functions \cite{Heun}. For
$\varepsilon\neq 0$ the functions
$R_{\,\,\varepsilon,s,l}^{(1)\pm}(r)$ (\ref{r1}) define a pair of
linearly independent solutions with the symmetry property
\ben R_{\,\,\varepsilon,s,l}^{(1)\pm}(r)=
R_{\,\,-\varepsilon,s,l}^{(1)\mp}(r).\la{Rsym}\een

\section{The Interior Spectral Problem}
\subsection{The Continuous Spectrum  ($\varepsilon^2>0$).}

\begin{figure}[htbp] \vspace{5.truecm}
\begin{center}
\includegraphics{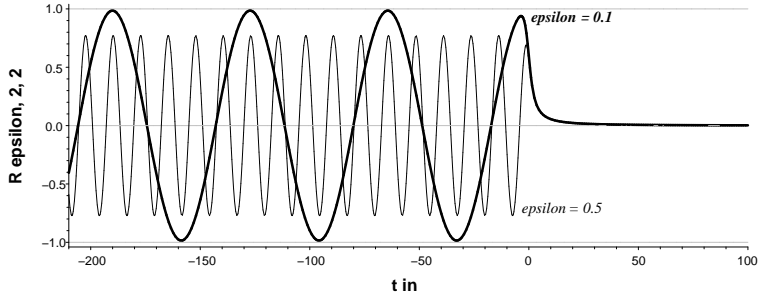}
    \vskip -1.truecm
\end{center} \caption{Two solutions of the continuous
spectrum.}
    \label{Fig0}
\end{figure}

For real $\varepsilon^2>0$ the problem has a real continuous
spectrum. The corresponding solutions $R_{\varepsilon, 2,
2}(t_{in})$ are illustrated in Fig. \ref{Fig0}. They define a basis
of stable {\em normal modes} for perturbations of the metric in the
SBH interior and can be used for Fourier expansion of perturbations
of more general form.

As seen in the Fig. \ref{Fig0}, the use of the inner time variable
$t_{in}$ instead of the tortoise coordinate $x$ makes the variations
of the solutions almost equidistant in the limit $t_{in}\to
-\infty$. This property makes the inner time variable $t_{in}$ to
seem the most natural one in the SBH interior. Using variables like
$r$ or $x$ we lose this property and meet numerical problems in this
limit.

\subsection{The Special Solutions with $\varepsilon^2=0$.}
The zero eigenvalue $\varepsilon=0$ yields a degenerate case in
which the two functions (\ref{r1}) coincide. Then we obtain the
infinite series of {\em polynomial} solutions:
\ben
R_{\,\,\varepsilon=0,s,l}^{(0)-}(r)\!=\!{{(l\!-\!s)!}\over{Poch(1\!+\!2s,l\!-\!s)}}
r^{s\!+\!1} JacobiP(l\!-\!s,2s,0,1\!-\!2r)=r^3\!+\!\dots,
\la{e0}\een
which are finite on the whole interval $r\in[0,1]$.  These functions
form an orthogonal basis with respect to the measure $w(r) =
(1-r)^{2s} $. Here $JacobiP(n,a,b,z)$ stands for the standard Jacobi
polynomials and $Poch(c,d)$ -- for the Pochhammer symbol.

The basis of the stable normal modes (\ref{e0}) is suitable for
series expansion of the perturbations of $\varepsilon= 0$. These
solutions do not depend on the variable $r_{in}$ and are the only
ones, which are finite at {\em both} singular ends of the interval
$r\in[0,1]$.

\subsection{The Discrete Spectrum ($\varepsilon^2<0$).}
The discrete spectrum of the problem at hand corresponds to pure
imaginary values of the parameter $\varepsilon$, i.e. to Laplace
transform of the perturbations of general form, instead of their
Fourier transform, which corresponds to real values of
$\varepsilon$. Thus, studying the the case $\varepsilon^2<0$ we are
constructing a basis for Laplace expansion of the perturbations of
more general form in the SBH interior.

For a qualitative analysis of the discrete spectrum we can apply the
Ferrari-Mashhoon transformation \cite{VFBM}: $x\to \tilde x=-ix$,
$r_s\to \tilde r_s=-ir_s$, $V\to \tilde V=-V$. It reduces our
spectral problem to a standard one: to find the bound states in the
inverted potential $ \tilde V$. For $s\geq 2$ the latter describes
two negative-valued potential wells i) a finite one, in which one
may have several negative eigenvalues, and ii) an infinite one, in
which one has infinite series of negative eigenvalues. In both cases
$\varepsilon^2\leq 0$. Hence, $Re\left(\varepsilon\right)\equiv 0$.
Then the functions (\ref{r0}), (\ref{r1}) and all their parameters
are real. In particular,
\ben\la{R01}
%\begin{array}{}
R_{\,\,\varepsilon,s,l}^{(0)-}(r)\!=\!
\Gamma_{1+}^{0-}(\varepsilon,s,l) R_{\,\,\varepsilon,s,l}^{(1)+}(r)+
\Gamma_{1-}^{0-}(\varepsilon,s,l) R_{\,\,\varepsilon,s,l}^{(1)-}(r)
\\ \nonumber
\hskip 1.65truecm\sim \cos\alpha\,R_{\,\,\varepsilon,s,l}^{(1)+}(r)+
\sin\alpha\,R_{\,\,\varepsilon,s,l}^{(1)-}(r) \hskip 2truecm
%\end{array}
%
\een
with real transition coefficients $\Gamma_{1-}^{0\pm}$ \cite{F2005}
and real mixing angle $\alpha$:
\ben \tan\alpha=\Gamma_{1-}^{0-}(\varepsilon,s,l)/
\Gamma_{1+}^{0-}(\varepsilon,s,l)\in \mathbb{R},
\,\,\,\hbox{for}\,\,\,\varepsilon=i\varepsilon_I.\hskip
.3truecm\la{alpha}\een

The physical meaning of the mixing angle $\alpha$ is obvious: since
the solutions $R_{\,\,\varepsilon,s,l}^{(1)+}(r)$ describe the
ingoing into the horizon waves and the solutions
$R_{\,\,\varepsilon,s,l}^{(1)-}(r)$ -- the outgoing from the horizon
waves \cite{F2005}, this angle describes the ratio of the amplitudes
of these waves in their mixture (\ref{R01}). The value $\alpha=0$
corresponds to absence of the outgoing waves, and the value
$\alpha=\pi/2$ describes the case without ingoing waves. Thus,
fixing the value of the mixing angle $\alpha$ we are defining
completely the physical problem under consideration.

When $r\to 0$, the asymptotic of the inverted potential is $\tilde
V\sim (1-s^2)r^{-4}$. Hence, we arrive at the well known from
quantum mechanics specific physical problem of "falling at the
centrum" -- the singular point $r=0$. Mathematically this means that
the corresponding differential operator on the interval $r\in [0,1]$
has a defect \cite{RS} and we need an additional physical conditions
to fix the eigenvalue problem. In our case such conditions are the
regularity condition of solutions at the singular point $r=0$ and
the fixation of the mixing angle $\alpha$, i.e. the justification of
the physical conditions at the second singular point -- the horizon
$r=1$. Thus, for any given value of $\alpha$ the Eq. (\ref{alpha})
presents a specific spectral equation for the eigenvalues.

Unfortunately, at present the transition coefficients
$\Gamma_{1\pm}^{0-}(\varepsilon,s,l)$ are not known explicitly
\cite{Heun}. Therefore we replace the spectral equation
(\ref{alpha}) with the equivalent one:
\ben \left|\begin{array}{cc} R_{\,\,\varepsilon,s,l}^{(0)-}(r_1) &
\cos\alpha\,R_{\,\,\varepsilon,s,l}^{(1)+}(r_1)\!+\! \sin\alpha\,
R_{\,\,\varepsilon,s,l}^{(1)-}(r_1)
\\ R_{\,\,\varepsilon,s,l}^{(0)-}(r_2)
&\cos\alpha\,R_{\,\,\varepsilon,s,l}^{(1)+}(r_2)\!+\! \sin\alpha\,
R_{\,\,\varepsilon,s,l}^{(1)-}(r_2)
\end{array}\right|=0,\hskip .6truecm\la{spectralEq}\een
where $r_{1,2}\in (0,1), r_1\neq r_2$ are two otherwise arbitrary
points. Because of the symmetry property (\ref{Rsym}), it is enough
to study only the solutions $\varepsilon_{n,s,l}$ of the equation
(\ref{spectralEq}) with positive
$\varepsilon_I=Im\left(\varepsilon_{n,s,l}\right)>0$.

\begin{figure}[htbp] \vspace{6.truecm}
\begin{center}
\includegraphics{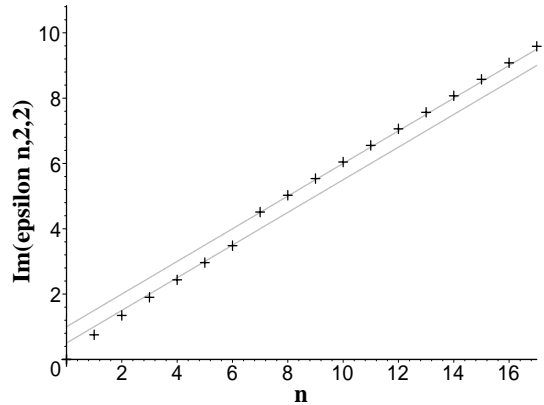}
\end{center} \caption{The first 18 eigenvalues (for $s=l=2$ and $\alpha=0$).} \label{Fig1}
\end{figure}

\begin{figure}[htbp] \vspace{6.truecm}
\begin{center}
\includegraphics{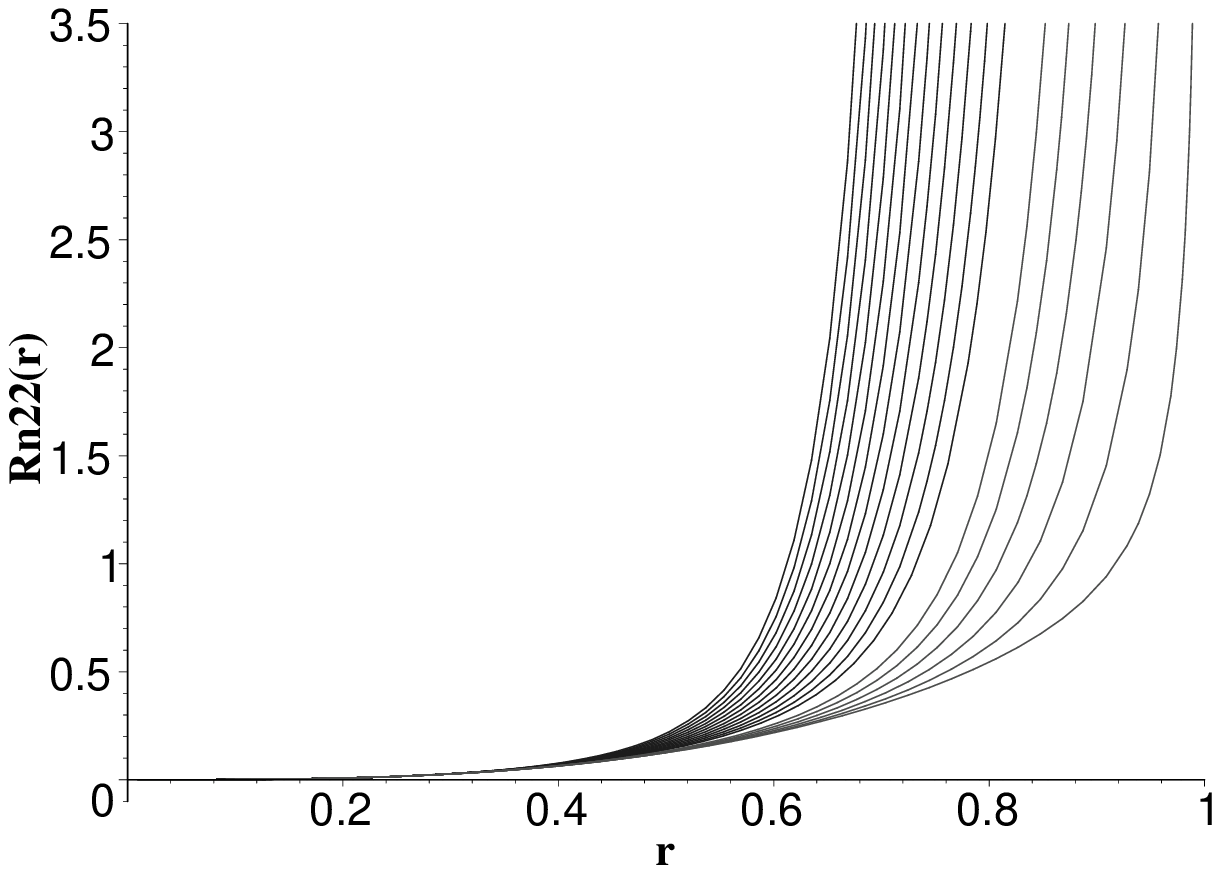}
\end{center} \caption{From right to the left: The first
18 eigenfunctions (for n=1, 2,\dots,18, s=l=2 and $\alpha=0$).
    \hskip 1truecm}
    \label{Fig2}
\end{figure}

\begin{figure}[htbp] \vspace{6.truecm}\begin{center}
\includegraphics{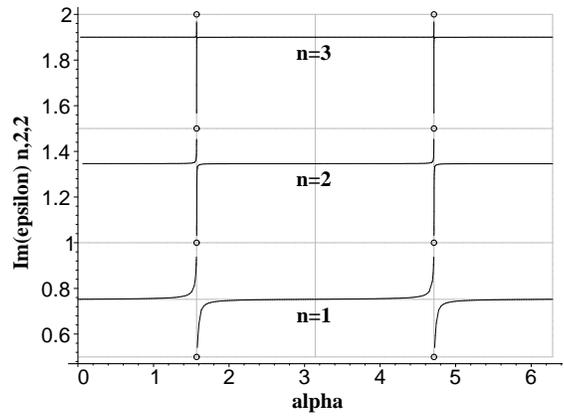}
\end{center}
\caption{The dependence of $Im\left(\varepsilon_{n,2,2}\right)$ on
the mixing angle $\alpha\in [0,2\pi]$.
    \hskip 1truecm}
    \label{Fig3}
\end{figure}

\begin{figure}[htbp] \vspace{6.truecm}
\includegraphics{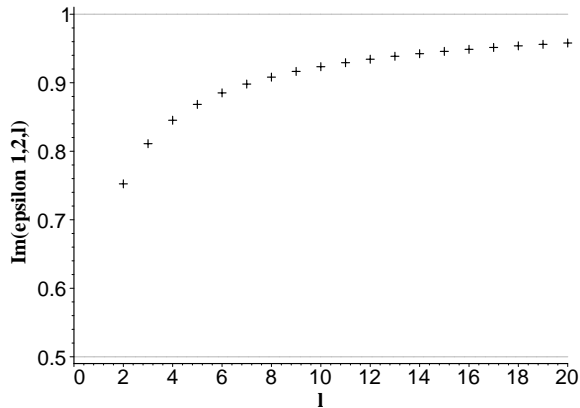} \caption{\hskip 0.2truecm The dependence of the
eigenvalues $\varepsilon_{1,2,l}(\alpha=0)$ on the angular momentum
$l$.
    \hskip 1truecm}
    \label{Fig4}
\end{figure}

We apply the complex plot abilities of Maple 10 to find a relatively
small vicinity of the complex roots of this equation and then we
justify their values by Muller's method for finding of complex roots
\cite{Muller}. Our numerical results for
$Im\left(\varepsilon_{n,2,2}\right)$ are shown in the Table
\ref{table1} and in  Figs. \ref{Fig1}--\ref{Fig4}.

\begin{table}[here]~\vspace{.truecm}
\begin{center}
\begin{tabular}{cllllrrrr}
\hline \hline
        $n$& $\Big|$ $Im\left(\varepsilon_{n,2,2}\right)$&
$\Big|$ $n$&$\Big|$ $Im\left(\varepsilon_{n,2,2}\right)$&
$\Big|$ $n$\hskip 0.25truecm $\Big|$ $Im\left(\varepsilon_{n,2,2}\right)$&\\
\hline
$1$&$\Big|$  0.7523& $\Big|$ $7$ & $\Big|$ 4.5138&  $\Big|$ $13$  $\Big|$  7.5657&\\
\hline
$2$&$\Big|$  1.3463& $\Big|$ $8$ & $\Big|$ 5.0256&  $\Big|$ $14$  $\Big|$  8.0714&\\
\hline
$3$&$\Big|$  1.8999& $\Big|$ $9$ & $\Big|$ 5.5357&  $\Big|$ $15$  $\Big|$  8.5768&\\
\hline
$4$&$\Big|$  2.4361&$\Big|$ $10$ & $\Big|$ 6.0446&  $\Big|$ $16$  $\Big|$  9.0823&\\
\hline
$5$&$\Big|$  2.9627&$\Big|$ $11$ & $\Big|$ 6.5524&  $\Big|$ $17$  $\Big|$  9.5867&\\
\hline
$6$&$\Big|$  3.4833&$\Big|$ $12$ & $\Big|$ 7.0594&  $\Big|$ $18$  $\Big|$  10.0939 &\\
\hline \hline
\end{tabular}
\end{center}
\caption{The values of the first 18 eigenvalues of the RWE for SBH
interior for $s=l=2$ and $\alpha=0$.} \la{table1}
\end{table}
\vskip 0.5truecm

In the Fig.2 we see the first 18 eigenvalues (including the zero
eigenvalue) of the RWE for SBH interior for $s=l=2$ and $\alpha=0$.
The two series: $n=0, 1,\dots6$; and $n=7,\dots$ are clearly seen.
They are placed around the straight lines $\varepsilon_I=n/2+1/2$
and $\varepsilon_I=n/2+1$, correspondingly.

In Fig.\ref{Fig3} we see a specific attraction and repulsion of the
eigenvalues $\varepsilon_{n,s,l}(\alpha)$. Such behavior is typical
for the eigenvalue problems related with the Heun's equation
\cite{Heun}. It is well known in the quantum physics. To the best of
our knowledge, this phenomenon is observed for the first time in the
theory of SBH.
Several additional remarks are in order:

1. Solving Eq. (\ref{spectralEq}) one is unable to obtain the
eigenvalues $\varepsilon_{N/2,2,2}({\pi\over 2})=i{N\over 2}, N\in
\mathbb{Z}^{+}$, i.e., when $\alpha\!=\!{\pi\over 2}/\!\!\!\mod \pi
$ (the small circles in Fig.\ref{Fig3}). A pure numerical study of
the RWE (\ref{R}) shows that in these cases the solutions have the
same behavior as the one, demonstated in Fig.\ref{Fig2}. Hence, one
has to be careful, taking the corresponding limits in the analytical
expressions (\ref{r0}), (\ref{r1}).

2. In the second series: $n=7,...$ the dependence of the eigenvalues
on the mixing angle is similar to the one, shown in Fig.\ref{Fig3},
but $Im\left(\varepsilon_{n,2,2}\right)$ are decreasing functions of
$\alpha$. The amplitudes of their variations are extremely small ($<
10^{-9}$) and they are practically constant on the larger part of
the interval $\alpha\in [0,2\pi]$.

3. There exist two more series of eigenvalues $\varepsilon_{n,2,2}$.
For them all $Im\left(\varepsilon_{n,2,2}\right)$ are negative.
These have the same absolute values as the eigenvalues, shown in
Table \ref{table1} and in Fig.\ref{Fig3}. Their dependence on the
mixing angle $\alpha$ is shifted by ${\pi\over2}$ and
correspondingly inverted. Because of this shift, the behavior of the
corresponding eigenfunctions is the same as the one, demonstrated in
Fig.\ref{Fig2}. This reflects the symmetry property (\ref{Rsym}) of
the solutions.

4. The behavior of the  eigenfunctions $R_{1,2,l}(r)$ with
$l=3,\dots$ is the same as of the functions $R_{1,2,1}(r)$ in
Fig.\ref{Fig2}. The dependence of the eigenvalues
$\varepsilon_{1,2,l}(\alpha=0)$ on the angular momentum $l$ is shown
in Fig.\ref{Fig4}.

5. The eigenfunctions $R_{n,2,l}(r)$ with $n>1, l>2$ start from
infinity at $r=1$ and approach zero at $r=0$, {\em oscillating}
around the zero value. The number of oscillations depends on the
numbers $n$ and $l$.

6. Our results bring to light a complementary information about the
stability problem of SBH. Up to now only the stability of the
exterior domain has been studied in linear approximation \cite{NM}.
Here we have found the following details about stability properties
of the solutions in the SBH interior:

$\bullet$ As we see in Figs. \ref{Fig0}, \ref{Fig2}, all solutions
of Eq. (\ref{RW}), which are considered in the present article, are
stable in direction of the interior-time-future ($t_{in}\to
+\infty$), {\em according to Lyapunov criterion}. If $\varepsilon
\neq 0$, these solutions are unstable in direction of the
interior-time-past ($t_{in}\to -\infty$) according to Lyapunov
criterion.

$\bullet$ There is a basic difference in the stability properties of
the solutions of continuous spectrum (real
$\varepsilon,\,\,\,\varepsilon^2
> 0$), which are of limited variation when $t_{in}\to -\infty$ and
thus -- stable in direction of the interior-time past in this weaker
sense, and the solutions of discrete spectrum (imaginary
$\varepsilon,\,\,\,\varepsilon^2 < 0$), which grow up infinitely
when $t_{in}\to -\infty$ and, hence, are unstable in this direction
of the interior time in any sense.

$\bullet$ All other interior solutions of the Regge-Wheeler equation
with $\varepsilon \neq 0$, which are not considered in this article,
are unstable in the future direction of the interior time and drop
out of the physical consideration.

$\bullet$ There do not exist interior solutions, different from the
above ones with $\varepsilon = 0$ (see Eq. (\ref{e0})), which are
stable in both directions of the interior time.

\section{Concluding Remarks}
In this article we have shown that the Heun's functions are an
adequate powerful tool for description of the linear perturbations
of the gravitational field not only outside the SBH horizon
\cite{F2005}, but in their interior domain as well.

It is well known that every analytical function is defined
completely by its singularities in the whole compactified complex
domain. In the previous investigations of the specific properties of
the solutions of RWE was used only the structure of these singular
points, see the review articles \cite{QNM} and the huge amount of
reverences therein. Using many different mathematical technics, this
way were reached important results for perturbations in the exterior
of SBH. Some of the used methods of that type may be not applicable
for the study of perturbations in the SBH interior.

Here we overcome the limits of that methods, using directly the
properties of Heun's functions, see for example Eq. (\ref{r0}),
(\ref{r1}), (\ref{Rsym}), (\ref{alpha}), etc. and writing down the
corresponding explicit analytical solutions of the RWE.

All numerical results were obtained using Maple 10 package for
performing {\em direct} calculations with these functions. Owing to
the new version 10 of the Maple package, at present we are able to
work with the Heun's functions in the complex domain, thus obtaining
a new freedom and new tools for the study of different new boundary
problems for RWE, as well as for study of the old ones, applying new
technics.

Using Heun's functions we have obtained the analytical and the
numerical description of all solutions of the Regge-Wheeler equation
in the SBH interior, which are stable in direction of the
interior-time future and single-valued functions in the whole
complex domain. Having in our disposal these solutions we are ready
for a more deep understanding of the problems of BH physics.

For example, as we have seen, the perturbations of SBH metric do not
change the singularity at the point $r=0$, at least in the first
order of perturbation theory. This result can be used for further
justification of the numerical treatment of BH problems without
excision of the interior domain. Now it  becomes clear, that due to
the unavoidable numerical errors in the interior of the BH one will
lose the small stable single-valued physical solution around the
point $r=0$ at the background of the big nonphysical solution.
Hence, in order to obtain physically correct and stable numerical
results one must suppress by proper techniques this nonphysical
solution.

We have shown that under proper boundary conditions there exist
interior solutions both of real continuous spectrum and of pure
imaginary discrete spectrum. The first may form a basis for Fourier
transform of linear perturbations of general form in SBH interior.
The second may form analogous basis for Laplace transform. The
investigation of the completeness of these sets of functions as a
basis for the corresponding expansions and the analytical proof of
the numerical results of the present article remain an open problem.

Our numerical study shows that there do not exist other solutions of
the spectral equation (\ref{spectralEq}). In particular, in the
two-singular-points-boundary problem on the interval $r\in[0,1]$ we
do not found solutions with $Re(\varepsilon)\times
Im(\varepsilon)\neq 0$, in contrast to the case of exterior domain
perturbations of SBH.

The discrete pure imaginary eigenvalues have a nontrivial behavior
under changes of the boundary conditions at the horizon, described
by the mixing angle $\alpha$. The phenomenon of attraction and
repulsion of these eigenvalues has been observed.

The basic properties of the considered solutions have been studied
and a new information about stability of the solutions of RWE in the
SBH interior was obtained.

In the present article we have studied only the most important from
physical point of view perturbations of SBH with $s=2$. The detailed
study of the solutions with $s=0, 1$ may be useful for numerical
relativity in the spirit of the article \cite{LBLR}.

\vskip .5truecm
{\em \bf Acknowledgments} \vskip .1truecm

The author is grateful to the High Energy Physics Division, ICTP,
Trieste, for the hospitality and for the nice working conditions
during his visit in the autumn of 2003. There the idea of the
present article was created. The author is grateful to the JINR,
Dubna, too, for the priority financial support of a cycle of
scientific researches and for the hospitality and the good working
conditions during his three months visits in 2003, 2004 and 2005.
This article also was supported by the Scientific Found of Sofia
University, Contract 70/2006, by its Foundation "Theoretical and
Computational Physics and Astrophysics" and by the Scientific Found
of the Bulgarian Ministry of Sciences and Education, Contract VUF
06/05.

The author is thankful to Nikolay Vitaniov for the reading of the
manuscript and the useful suggestions, to Luciano Rezzolla for the
useful and stimulating discussion of the results of the present
article and their relation with the recently found numerical
techniques for treatment of BH problems without excision of the
interior domain, and to Kostas Kokkotas for discussion of the exact
solutions of RWE and SBH interior solutions during the XXIV Spanish
Relativity Meeting, E.R.E. 2006.

%\vskip 1.truecm

\end{document}